\documentstyle[epsfig]{article}
\newcommand{\beq}{\begin{equation}}
\newcommand{\eeq}{\end{equation}}

\title{SEARCH FOR POSSIBLE VARIATION OF THE FINE STRUCTURE CONSTANT}

\author {Savely G. Karshenboim\thanks{E--mail: sek@mpq.mpg.de}\\
\medskip
{}\\ D. I.  Mendeleev Institute for Metrology (VNIIM),\\
St. Petersburg 198005, Russia\\
Max-Planck-Institut f\"ur Quantenoptik,\\
85748 Garching, Germany
}

\begin{document}

\maketitle

\begin{abstract}
Determination of the fine structure constant $\alpha$ and search for its 
possible variation are considered. We focus on a role of the fine structure 
constant in modern physics and discuss precision tests of quantum 
electrodynamics. Different methods of a search for possible variations of 
fundamental constants are compared and those related to optical measurements 
are considered in detail.
\end{abstract}

\section{Introduction \label{s:int}}

Fundamental constants play important role in modern physics. Some of them 
are really universal and enter equations from different subfields. An example 
is the fine structure constant $\alpha$ and other quantities related to 
properties of electron and nucleon. 
Electron is a carrier of the electric charge and it is the orbiting particle 
in any atom. Thus, its characteristics enter a great number of basic equations 
of atomic physics as well as various effects of electrodynamics, which are 
related to quantum nature of physics or discrete (atomic) nature of matter. 
The latter is a back door through which quantum physics enters nearly 
classical experiments involving the Avogadro number, Faraday constant and 
various properties of single species. The ``hidden" quantum mechanics is in 
the fact of the very existence of identical objects. Thus, the electron charge 
$e$ is a mark of an open or hidden quantum nature of a phenomenon. 

Most quantities related to properties of electron are dimensional and thus 
depend on our definitions of units. The fine structure constant, which in 
fact is the squared electric charge of electron in natural units 
\beq
\alpha=\frac{e^2}{4\pi \epsilon_0 \hbar c}\;,
\eeq
is dimensionless and thus its value is a ``true'' fundamental constant which 
does not depend on any convensions (like, e.g., definitions of units in 
International System SI) and it is one of very few fundamental constants 
which can be measured in actual physical experiments and originate from 
physics of very short distances related to Grand Unification Theories.

Proton and nuclei formed by protons and neutrons also transport electric 
charge in some phenomena and they are chiefly responsible for mass of 
classical massive objects. As attractors for orbiting electrons,
nuclei are responsible for recoil and hyperfine effects. They also offer 
some dimensionless parameters such as mass ratio of electron and proton 
$m_e/m_p$ and $g$ factors of free proton and neutron. However, such parameters 
are less fundamental than the fine structure constant.

A variety of completely different approaches to an accurate determination 
of the fine structure constant $\alpha$ shows how fundamental and universal 
this constant is. The value of $\alpha$ can be determined from quantum 
electrodynamics (QED) of a free electron, from spectroscopy of simple atoms, 
from quantum nonrelativistic physics of atoms and particles, from measurement 
of macroscopic quantum effects with help of electrical standards. A comparison of 
different approaches to the determination of the fine structure constant 
allows us to check consistency of data from all these fields, where the 
determination of $\alpha$ involves the most advanced methods and technologies. 
The further output of this activity is precision QED tests and the new 
generation of electrical standards.

A possibility of time and space variations of some fundamental constants is 
very likely for a number of reasons. For example, we accept a so-called 
inflatory model of the evolution of our Universe and thus have to acknowledge 
a dramatic variation of the electron mass and electron charge at a very early 
moment of the evolution of the Universe. Therefore we could expect some much 
slower variations of these and some other quantities at the present time. A 
possible variation of fundamental constants is searched for in two kinds of 
experiments:
\begin{itemize}
\item measurements of the quantities easily accessible and/or potentially 
highly sensitive to the variations (such as experiments at the Oklo reactor);
\item studies of the quantities that allow a clear interpretation, particularly
in terms of fundamental physical constants.
\end{itemize}

Indeed, the only fact of a cosmological variation detected today and 
determination of its magnitude would be a great discovery and one could think 
that its clear interpretation is not so important at the initial stage of a 
search.

However, everybody experienced in high precision studies knows that while 
achieving a new level of accuracy, one very often encounters new systematic 
effects and sometimes it takes quite a while to eliminate them. We can expect 
receiving a number of different positive and negative results and without a 
proper interpretation we will hardly be able to compare them and check their 
consistency. Actually, such results have already been obtained:
\begin{itemize}
\item Studies of samarium abundance in uranium deposits in Oklo showed 
{\em no significant shift} of the 97.3 meV resonance while comparing the 
current situation with the operation of the Fossil natural fusion reactor 
about 2 billion years ago \cite{schlyachter}. Since the typical nuclear 
energy is a few MeV per nucleon, that is a strong statement. We remind that 
meV stands for $10^{-3}$ while MeV for $10^6$ eV.
\item Studies of the quasar absorption spectrum imply {\em a relative shift} 
of atomic lines at the ppm level. That is a line-dependent shift (additional 
to the common Doppler red shift) related to the period of up to ten billion 
years \cite{webb}. It may be interpreted as an evidence that a value of the fine structure constant was in the far past not the saem as now.
\item Laboratory comparison of hyperfine splitting in cesium and rubidium 
finds {\em no variation} at the level of a few parts in $10^{15}$ per 
year \cite{salomon}.
\end{itemize}

With two negative and one positive results we still cannot make a statement 
because they are not comparable. We need some set of data obtained in a really 
independent way but with results clearly correlated in the case of the 
variation of constants.

Since quantum electrodynamics is the best advanced quantum theory and the 
fine structure constant is a basic electrodynamical quantity, it is attractive 
to use it as an interface for a search for a possible variation of 
constants. Such an interface allows clear interpretation and a reasonable 
comparison of different laboratory experiments, which will hopefully deliver 
their results in one or two years.

The need to interpret the results in terms of some fundamental parameters 
comes basically from two reasons:
\begin{itemize}
\item It is of crucial importance to be able to compare results on the 
variation of various quantities coming from different experiments. To be 
compared, they have to be expressed in terms of the same universal parameters.
\item Theory of Grand Unification and quantum gravity can be helpful to 
establish links between the variation of different fundamental constants, 
such as $\alpha$ and $m_e/m_p$ etc. The possible links can also extend to 
experiments on a search for other {\em exotic} effects such as a violation 
of the equivalence principle. 
\end{itemize}

The paper is substantially based on a talk at the HYPER symposium (Paris, 
2002). Since a number of important results such as \cite{kinoshita,ivanchik,marion,wilpers,bize,fischer,peik,lea} were published just after the meeting, the 
paper has been considerably enlarged and updated for this publication.

\section{Quantum electrodynamics and the fine structure constant \label{s:qed}}

Quantum electrodynamics (QED) of free particles and bound atomic systems 
allows a number of accurate tests. Most comparisons of theory versus 
experiment mainly confirm theory, while some of experiments show either 
data scatter or slight disagreement within a few standard deviations which in principle 
can be expected due to a big number of different experimental QED tests.
\begin{itemize}
\item Free QED can be tested via a study of the anomalous magnetic moment of 
electron and muon. The former is limited by our knowledge of the fine 
structure constant $\alpha$ while for the latter the dominant sources of the 
uncertainty are the experimental uncertainty and inaccuracy of our 
understanding of hadronic effects, i.e. effects of strong interactions.
\item Bound state QED can be tested for weakly bound atoms (when the binding 
energy in low-$Z$ atoms that is proportional to $(Z\alpha)^2mc^2$, is 
significantly smaller than $mc^2$) or for the strong coupling regime realized 
in high-$Z$ few-electron atoms. Advanced QED calculations allow to reach such 
a high accuracy that for any values of the nuclear charge $Z$ the theoretical 
uncertainty for conventional atoms is due to the nuclear-structure effects. 
The experimental uncertainty and the uncertainty of the determination of 
fundamental constants are sometimes not so small and for any QED tests 
performed with hydrogen-like atoms, the uncertainty of the pure QED 
calculations is not a limiting factor today \cite{sgk_ap,sgk_2003}.
\end{itemize}

\begin{figure}
\centerline{\epsfig{figure=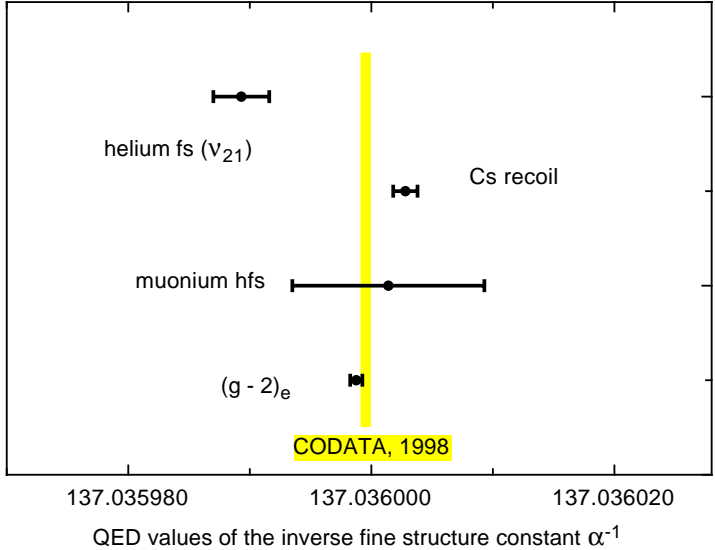,height=3.0in}}
\caption{Determination of the fine structure constant $\alpha$ by means of 
QED and atomic physics. The presented values are from the anomalous magnetic 
moment of electron \protect\cite{vandyck} (recently corrected in 
\protect\cite{kinoshita}), the muonium hyperfine structure \protect\cite{liu}, 
the helium fine structure \protect\cite{drake,george} and the 
recoil-photon experiment with cesium \protect\cite{wicht} (the 
proton-to-electron mass ratio is taken from \protect\cite{beier,km_plb}). The 
CODATA result is related to the adjustment performed in 1998 
\protect\cite{mohr}.\label{Fig:alpha}}
\end{figure}

To perform a successful QED calculation, one needs to know proper values 
of several fundamental constants related to atomic physics and quantum 
electrodynamics. One of the most important is the fine structure constant. 
Its determination by means of atomic physics and QED is summarized in 
Fig.~\ref{Fig:alpha}.

We have not included into Fig.~\ref{Fig:alpha} any results for which the 
uncertainty is mainly due to electrical and material standards since the 
basic strategy is to determine $\alpha$ without involving those standards 
and apply its value to calibrate the standards. An accurate value of the 
fine structure constant is of great practical importance for electrical 
standards, mainly due to reproduction of {\em ohm} \cite{taylor1} and, 
perhaps in future, of {\em kilogram} \cite{taylor2}. The same value of 
electrical resistance in any measurement performed at different laboratories 
can be maintained with the help of the quantum Hall effect without any 
knowledge of the fine structure constant. However, an accurate value of 
$\alpha$ is still needed for a proper {\em reproduction} of the basic 
electrical units of SI in order to avoid any inconsistency between the units 
of different physical quantities. For example, the mentioned above quantum 
Hall resistance (so called von Klitzing constant) is
\begin{equation}
R_K = \frac{h}{e^2}=\frac{\mu_0c}{2\alpha}
\;.\end{equation}
If the electrical standards accomodate an incorrect value of the quantum Hall 
resistanse $R_K$, it will lead to either a breakdown of the Ohm law (if  
{\em ohm}, {\em volt} and {\em ampere} are treated in an inconsistent way) or 
to a discrepancy (via the Ampere law) between {\em watt} determined from 
electrical and mechanical units. 

\section{Are fundamental constant really fundamental?\label{s:fund}}

Theory is not in position to produce any quantitative prediction, giving 
instead some expressions for the quantities that can be measured. To obtain 
any numerical results for those theoretical expressions, one needs some 
values of the fundamental constants for the input data. This need for proper 
input data limits accuracy even in a perfect theory. Those fundamental 
constants originate from the Hamiltonian that describes free particles and 
their interactions.

However, one needs to distinguish clearly between the unperturbed and 
perturbed parameters of Hamiltonian.
\begin{itemize}
\item The former are real fundamental constants, which in some scenario may 
be even fixed and in principle can be calculated. However, the unperturbed 
``bare'' parameters, such as a bare electron charge $e_0$, have 
{\em no direct relation} to any actual measurements.
\item In contrast, the latter are the constants we only see in our experiments 
as a measured electric charge, mass etc. They are a result of perturbations 
and renormalizations and from the fundamental point of view they are only 
{\em effective} parameters, like, e.g., effective mass of an electron in 
some medium. A difference between condensed matter physics and particle 
physics is that an electron can be extracted from any medium and studied as 
free, while in the case of particle physics we cannot study an electron free 
of QED interactions.  
\end{itemize}

The details of the renormalizazion $\delta e = e-e_0$ depend on physics at 
extremely short distances for which we have neither a proper theory nor 
experimental data. The theory we deal with is not appropriate there. However, 
for conventional problems we used not to care about bare parameters targeting 
to bridge different measurements and thus expressing measured cross sections, 
lifetimes and energy shifts in terms of measured masses and charges. This 
idea was a breakthrough at the early time of QED and it opens a way to a 
selfconsistent description of physics of our ``low-energy" world. To work for QED one 
needs to deal only with measurable quantities and not to care about their  
``real" origin. However, a situation becomes quite different when we turn to 
the consideration of the variation of fundamental constants. In such a case 
we have to care about the origin of both the bare charge $e_0$ and the 
renormalization correction $\delta e$ and in particular about its dependence 
on details of physics at the Unification and Planck scale 
($M_{\rm Pl}=(c\hbar/G)^{1/2}\simeq 1.2\cdot 10^{19}\;{\rm Gev}/c^2\simeq 2.2\cdot 10^{-8}\;{\rm g}$).

Thus, we arrive to a strange situation when there are some undetectable 
truly fundamental constants and some universal experimental values related 
to perturbed effective parameters.

Discussing a status of different fundamental constants, we have to note that 
there is some hierarchy among them.
\begin{itemize}
\item The most fundamental constants are constants related to properties of 
space-time, namely, $\hbar$, $c$ and $G$, which determine the Planck units.
\item Another important set of fundamental constants is formed by those which are 
responsible for properties of basic particles:
 \begin{itemize}
 \item electromagnetic, weak and strong coupling constants (and, in 
particular, the electric charge $e_0$);
 \item $\Lambda_{\rm QDC}$, a dimensional parameter of quantum chromodynamics, 
which mainly determines observed mass of proton and neutron;
 \item Yukawa coupling constants of interaction of ``normal" particles with 
the Higgs particles which determine the lepton, quark masses and masses of 
intermediate bosons;
 \item parameters of the Higgs Lagrangian;
 \item parameters of the Cabibbo-Kobayashi-Maskawa matrix. 
 \end{itemize}
\end{itemize}
One can see that most of them cannot be measured precisely while some cannot 
be measured at all. However, it may be useful to remember about them when 
discussing a scenario of hierarchy of  $\alpha$, $m_e$ and $m_p$ variation. Other constants such as proton mass or magnetic moment are less fundamental.

\section{Are the fundamental constant really constants?\label{s:const}}

\subsection{Expressing results in terms of basic constants\label{s:exp}}

Since we distinguish between perturbed and unperturbed parameters of Hamiltonian, the most obvious scenario is 
when variations are related to effective perturbed constants, while the truly 
fundamental constants do not vary. However, a direct variation of parameters 
of the basic Hamiltonian may be also possible.

In the introduction we made a statement that it is crucially important to be 
able to compare different results. Some measurable quantities can be easily 
related to the fundamental constants. 
\begin{itemize}
\item Optical transitions can be in principle expressed in terms of the 
Rydberg constant and the fine structure constant.
\item A calculation of the hyperfine intervals requires some knowledge of 
the magnetic moment of nuclei. A model that presents all nuclear magnetic moments in 
terms of a few basic constants is necessary. An only model, which can express 
the nuclear moments in simple terms {\em ab initio}, is the {\em Schmidt model}
(see Sect.~\ref{s:hfs}). Unfortunately, it is not accurate enough and 
introduces an uncertainty that is sometimes large and probably cannot 
be reduced.
\end{itemize}

There are a number of values which cannot be expressed in simple terms mainly because nuclear 
effects dominate are strongly involved there. One example is a low-lying resonance in samarium, similar 
situation arises with a study of radioactive nuclear decay in samples from 
early time of either Earth or meteorites, investigations of pulsar periods 
{\em etc}.

\subsection{Basic constants\label{s:basic}}

We do not discuss here any limits for a possible variation of the 
gravitational constant $G$ because they are much weaker than limits from 
spectroscopy. We focus our consideration on atomic and in part molecular 
spectra, since they can be studied with high accuracy and can be expressed in 
terms of some basic constants.

From {\em the experimental point of view} the basic constants for 
interpretation of spectroscopic data are
\begin{itemize}
\item the fine structure constant $\alpha$;
\item proton-to-electron mass ratio $m_e/m_p$;
\item $g$ factor of proton;
\item $g$ factor of neutron.
\end{itemize}

The Rydberg constant $Ry$ is not included because its variation cannot be 
detected. Any results can be obtained only for {\em dimensionless} quantities 
which are mainly related to the frequency/energy ratio.
Rigorously speaking, a search for a variation of a dimensional value is also 
possible but another kind of experiment is needed. All experiments for a 
search of space and time variation of fundamental constants suggest two 
spectroscopic measurements separated in time and space. In the case of the 
astrophysical spectroscopic data evaluation, the {\em drift} of the Rydberg 
constant, if it happens, cannot be separated from the {\em red shift} due 
to {\em expansion} of the Universe or a peculiar motion of the spectroscopic and is to be 
interpreted as its part. To detect a drift of 
dimensional values  in laboratory experiments, we should check that the 
units are the same. In practice this involves another measurement and thus 
the signal for variation will be related to some ratio of their data, i.e. 
it should be a relative measurement.

However, a measurement of time and/or space variation of a dimensional 
quantity is still possible if an experiment is designed to deal directly with 
gradients. Doing spectroscopy we can look for a variation of the fine 
structure constant which is dimensionless. However, looking at the propagation 
of light we could judge on variability of the speed of light $c$ which is 
dimensional. It seems that a search for gradients is much more complicated.

In the case of the Rydberg constant it would be of interest to compare  
radiation from the same sources separated in time. However, 
in a laboratory experiment there is no reasonable way to delay the 
radiation for a long enough period and thus to compare a frequency of light 
sent from the same sources at different times. If a proper scheme is  
suggested, it should be applied first to detect the red shift due to 
the expansion of the Universe which is at a level of a part in $10^{10}$ per 
year, while the variation of constants should be at least a few orders of 
magnitude lower as we can see from various experiments \cite{schlyachter,webb,salomon,ivanchik} (see also Sect.~\ref{s:sum}). 

One can see that the above mentioned set of ``practical" fundamental constants is really 
different from the ``truly" fundamental one discussed above. This difference 
is really important 
from {\em the theoretical point of view}. In particular, looking for really 
fundamental constants instead of the proton-to-electron mass ratio, one has 
to deal with the electron and the proton mass. They are to be expressed in 
natural units of Grand Unification Theories (GUTs) which are related to either 
the Planck mass or the Unification mass. Since the origin of the electron mass 
(via the Higgs sector) and the proton mass (via QCD confinement) is different, 
they have to enter any theoretical scenario {\em separately}.

Expressing masses in units of the Planck mass we arrive to an experimental 
problem. The Planck mass enters experimental data only via gravitation effects.
However, there is no data for that (see, e.g., the review \cite{uzan}) with 
spectroscopic accuracy. An exception might be a study of pulsar periods. 
Potentially they can provide us with more accurate data. However, there is a 
number of reasons for the red shift of their period and the pulsar data can be 
rather of use if the variation of constants leads to a {\em blue shift} of 
the pulsar period. However, even in this case there are at least two sources 
of the blue shift:
\begin{itemize}
\item an increase of the mass of the pulsar because of picking up some 
amount of dust particles;
\item a decrease of the Hubble red shift if the pulsar is going towards the
Earth. 
\end{itemize}
The red shift is proportional to the distance between the emitter (the pulsar) 
and the absorber (the Earth) and it has to decrease when they move towards 
each other. The value of the decrease is
\beq
\Delta z= - H v \Delta t\;, 
\eeq
where $H$ is the Hubble constant and $v$ is the relative velocity. For 
$v = 200$~km/s the decrease of frequency is $3\times10^{-14}$ per year. 
In order to interpret the blue shift of the pulsar period at a level of a 
part per $10^{15}$ per  year, one has to determine the radial component of 
the pulsar velocity  with the uncertainty below 10 km/s. Meanwhile, the 
radial component of velocity cannot be determined so accurately. It is 
unlikely that an accuracy with which a period of any puslar can be determined 
is good enough to compete with spectroscopic data, however, they can deliver 
accurate limits on a variation of the gravitational constant $G$, compatible 
with other limitations.

\subsection{Correlations between variations of different 
constants\label{s:corr}}

There are two main problems related to the correlation of different basic 
constants. One is due to relations between truly fundamental constants and 
observable basic parameters and the other is due to relations between different 
fundamental constants. Let us start with the first one.

From the theory of strong interactions we know that the proton and neutron 
mass in the so-called chiral limit as well as any other dimensional quantities 
related to strong interactions are completely determined by a single parameter 
$\Lambda_{\rm QCD}$. The dimensionless quantities are just constants. However, 
quarks possess masses. Another violation of the chiral limit occurs due to 
electromagnetic interactions which involve leptonic loops. In other words,
the renormalized electric charge of quarks and nucleons depends on electron 
mass. Let us discuss dependence on quark masses in more detail. It is well 
known that a parameter $m_q/\Lambda_{\rm QCD}$ is small for $u$ and $d$ quarks
only, while it is not small for heavier quarks such as $s$ and $c$. We remind 
that light quark masses are below 10 Mev/c$^2$, while 
$\Lambda_{\rm QCD}=$216(25) Mev/c$^2$ and a mass of the $s$ quark is 
between 80 and 155 Mev/c$^2$ \cite{pdg}.

A fraction of heavier quarks is nearly vanishing in proton and neutron 
and does not contribute to some of their properties, such as magnetic moment. 
Effects of the light quark masses can be taken into account via chiral 
perturbation theory (see \cite{flam_shur} and references therein). However, 
heavy quarks contribute to the proton and neutron mass at the level of 
10-20\% because their contributions are enhanced by their masses and such 
contributions cannot be systematically taken into account. Note that 
the dependence of their contribution on the masses is not simple because the 
fraction of $s\overline{s}$ and $c\overline{c}$ pairs inside a proton depends 
on their masses.

Due to that, a practical question arises about fixing a set of basic 
``practical" constants in the form as above. In particular, one can suggest 
to combine $g$ factors of nucleons and the mass ratio $m_e/m_p$. A reason 
for that is their sensitivity separately to the contents of the strange quark, 
while their combination such as $\mu_p/\mu_B$ does not depend on that. 
However, the nuclear magnetic moment has the orbital component (proportional 
to the nuclear magneton) and spin component (proportional to the 
proton/neutron magnetic moment). The magnetic moment of a nucleon (proton or 
neutron) does not depend on effects due to $s$ and $c$ quarks, while the 
nuclear magneton determined via the proton mass does
\beq
\mu_{\rm N} = \frac{e\hbar}{2m_p}\;.
\eeq
Thus, magnetic moments and $g$ factors actually enter equations in different 
combinations some of which depend on the $s$ quark mass, while some not. A 
choice of a basic set is rather a matter of taste, however, variation of mass, magnetic moments and $g$ factors are correlated.

Variations of all fundamental constants are correlated and caused by the same 
reasons and in principle the rate should be approximately the same. However, 
some hierarchy is still in place and it depends on a scenario. Let us present 
some examples.
\begin{itemize}
\item The phase transition in the inflatory model happened because of general 
cooling, caused by the expansion, and temperature dependence of the vacuum 
expectation of the Higgs field. In that case the biggest rate corresponds to 
the electron mass. QED with massless charged leptons and in particular charge 
renormalization effects in such a theory differ very much from 
``conventional" QED with massive charged particles. In the massless case the 
electric charge does not exist in ``conventional" sense since it is defined 
as a coupling constant for the interaction at the asymptotically large 
distance. The result of such a definition is divergent. However, that  
does not matter indeed in the hot and dense Universe where there is some 
characteristic distance between particles. After the phase transition which 
provides leptons and quarks with their masses, the Universe is still 
expanding and cooling and the electric charge varies now with time and 
temperature via the vacuum polarization contributions to renormalization but 
at a negligible and undetectable level. The variation of the electron mass 
dominates. The variation of the electric charge of the electron is slower 
than that of the electron mass. The variation of the proton mass is mainly 
caused by the contents of the strange quarks; the same is with $g$ factors 
of the proton and neutron. The variation of the proton and neutron 
magnetic moments is even smaller and of the order of $g_p/g_n$. 
\item In the case of some oscillations (within the cosmological time scale) 
of the compactification radius, which enters as an effective cut-off of 
ultraviolet logarithmic divergences for the renormalization of the mass and 
charge of particles and likely determines parameters of the interaction 
with the Higgs sector and $\Lambda_{\rm QCD}$, one should expect relatively 
fast variations of the electron and proton masses separately while the 
variation of their ratio strongly depends on details of the model and in 
principle can vanish at all. The variation of the fine structure constant is 
to be slower than the variation of the proton mass \cite{marciano,calmet}. 
The variation of the proton $g$ factor is to be proportional to the strange 
part of the proton mass and a rate for the proton-to-electron mass ratio. 
The same is true for the neutron $g$ factor.
\item One more scenario can be related to a kind of a domain structure for 
some parameters which are not directly coupled to the vacuum energy, like,
e.g., the parameters of the Cabibbo-Kobayashi-Maskawa (CKM) matrix and in 
particular the Cabibbo angle. In such a case the biggest variation related to 
spectroscopy is to be for the $g$ factor of the proton and neutron. A domain 
structure may imply some relaxation on the way to the homogenous equilibrium, which has to 
produce a time variation to follow space variations. Another similar 
possibility is due to a variation of some mass ratio such as $m_e/m_\mu$  
when the vacuum energy is not changed. However, the most important effect is 
not directly related to spectroscopy. A shift of parameters of the CKM matrix 
or mass ratios will strongly affect the nuclear synthesis chain and it may 
happen that the same isotopes as on the Earth exist but with a 
different abundance, or even some isotopes stable here are unstable with 
shifted constants and vice versa. The spectroscopic study of transitions in the gross and fine structure from astrophysical sources cannot resolve different isotopes of the same element and with different isotopes presented the spectroscopic 
data will be indeed affected. Since relatively few elements are involved 
in analysis, the $\alpha$ variations may be emulated by a proper correction 
of fundamental constants which determine the nuclear synthesis. Lack of an accurate theory does not allow to eliminate the competition between possible $\alpha$ variation and different nuclear systhesis. This question requires further studies.
\end{itemize}

\section{Hierarchy and scenario: two examples}

A possibility that values of the fundamental constants vary with time at a 
cosmological scale was first suggested quite long ago \cite{dirac}, but no 
natural idea about a reason for such a variation has been suggested since. There are a number of simple reasons related to the evolution of Universe (cooling, expansion etc) which imply some small slow variation but not at a detectable level.
Instead, there is a number of different models and ideas about what can cause 
the some bigger variations (see, e.g., \cite{uzan}) which can in principle be detected present days, but the truth is that we do not 
really understand quantitively the origin of a number of basic quantities (fundamental constants) of our world and 
in particular cannot understand details of their possible variation in time and space.

Let us consider some of the examples above in more detail. We assume that 
most parameters may be in principle calculated and derived within some Grand Unification Theory. One may note 
that $\alpha\sim 10^{-2}$ and $m_e/M_{\rm Pl}\sim 4\cdot 10^{-23}$, so that 
these quantities are of such a different order of magnitude that it is 
hard to expect that both are calculated on the same ground. However, they 
can indeed become of the same order if we compare $\alpha$ and 
$\ln(M_{\rm Pl}/m_e)\simeq 51.5$. If they could be derived from first 
principles, the result must be something like $1/\pi^4$ or $\pi^{-2}/4\pi$ and 
corrected due to renormalization. How can the variations appear in such a 
scheme? We have two basic suggestions.
\begin{itemize}
\item The values derived {\em ab initio} are related to some equilibrum state. 
However, in a given moment $t$ we observe a compactification radius $R(t)$ 
oscillating at a cosmological scale around an equilibrium value $R_0$, which 
is completely determined by the Planck length $L_{\rm Pl}=\hbar/M_{\rm Pl}c$. 
The observable fundamental constants depend on $R(t)$ via their 
remormalization and thus vary. Most of the consequences were considered above 
in the previous section.
\item Some parameters are chosen due to a spontaneous violation of symmetry. 
They may have space variation (especially if this breakdown happened at 
the late period of the inflation) and their scatter will cause the space 
variation while their relaxation may be responsible for the time variation.
\end{itemize}  

The inflatory model assumes some phase transition with a breakdown of 
symmetry and fast expansion at the same time. This phase transition was 
suggested (see, e.g., \cite{inflation}) because we have to explain why 
essential properties of different parts of our Universe are the same with 
those parts in an earlier period although they are separated by horizon and 
cannot interfere with each other. The problem was solved by suggesting the 
inflation phase: a phase transition with so fast expansion that two points 
which were close to each other after the transition become separated by a 
horizon for a long cosmological time period. As we discussed in \cite{sgk_tam},
it may produce some space non-homogenous distribution for some parameters 
just after the inflation and their independent time evolution in the remote 
horizon-separated area with a cosmological scale. 

One can note that we have a number of parameters, the origin of which we do 
not undestand (see, e.g., \cite{pdg}):
\begin{itemize}
\item parameters of the Cabibbo-Kobayashi-Maskawa (CKM) matrix, which are 
responsible for different effects of the weak interactions with quarks and 
hadrons;
\item the ratio $m_\mu/m_e$ and other mass ratios for leptons and quarks 
from {\em different} generations. Actually, even within the same generation we cannot explain values of $m_e/m_u$ and $m_d/m_u$. 
\end{itemize}
Let us assume that they are a result of the same spontaneous breakdown of 
symmetry, which caused the inflation, and their values are determined 
worldwide during the inflation phase. We also suggest that they are not 
directly coupled to the vacuum energy\footnote{There is a number of parameters 
due to the CKM matrix and mass ratios and we can consider the vacuum energy 
as their function. The energy definitely depends on some parameters, however, 
it may happen that there is a valley in some direction of the multiparameter 
space. In other words, we suggest that it is possible to coherently change 
some parameters in such a way that the vacuum energy is not changed.} so 
that their space fluctuations cannot have the same value independently. 
There is no reason for them to be exactly the same everywhere and thus they 
can slightly vary in space and time and a pattern of their current values 
should be ``frozen out'' with an increase of the horizon separating 
different areas of the Universe. The space dependence will lead to time 
variations on a way to the equalibrium.

Let us briefly discuss the consequences.
\begin{itemize}
\item In the leading approximation, a weak-interaction effects related to the CKM matrix will lead to some correction to the magnetic moment of proton and neutron only at a level of 
a part in $10^5$ and thence the variation of the magnetic moments should be $10^5$ weaker than a variation of the CKM parameters. The proton and electron mass will not be shifted as well as 
the fine structure constant. In a cosmological scale they will also contribute 
to parameters of effective weak interactions and in particular to neutron lifetime.
\item The variation of mass parameters related to different generations like,
e.g., $m_\mu/m_e$ has unclear consequences for the fine structure constant 
because we do not really know the origin of the ratio $m_\mu/m_e$ and do not know which combination of $m_e$ and $m_\mu$ is coupled to the vacuum energy. Thus we 
cannot guess what a variation by, e.g., a factor of two means for either of two masses 
(in natural units of the Planck mass). It may happen that the electron and 
muon masses will be changed by two orders separately, or on the contrary, 
one mass can go up and the other down. In the first case a contribution to 
renormalization will be strongly affected. If masses themselves vary much 
faster than their ratio, a variation of the fine structure constant will be 
compatible with the variation of the mass ratio. In the second case the 
renormalization is weaker because the variation of the contributions of 
electron and muon loops will in part compensate each other and because they 
will be also suppressed by logarithmic nature of renormalization. The magnetic 
moments of the proton and neutron will be not changed. The proton mass will 
vary slightly because of the contribution of the current quark masses
(particularly of that of the $s$ quark), while the electron mass can vary 
with any rate. We see that any clarification of details is only possible
after a model is specified.\\ 
It is most likely that the lepton and quark masses should be somehow coupled 
to the vacuum energy. But some their specific combinations could still be 
uncoupled (see the footnote above). It is more natural to expect that vacuum energy should depend on some ``averadge" mass of all leptons and quarks and thus the second scenario is more probable. 
\end{itemize}
The hierarchy is very different from, e.g., the one suggested 
in \cite{calmet,langacker} with a much faster variation of $m_p/m_e$ than 
that of the fine structure constant $\alpha$ and the $g$ factors. 
We discuss these two examples in order to demonstrate that hierarchy 
depends on a scenario and there is a great number of very different options 
for them. Further discussion of the hierarhy and scenaria is presented 
in \cite{sgk_tam}.

\section{Atomic spectroscopy and variation of fundamental 
constants\label{s:at}}

High-resolution spectroscopy offers a possibility to study variations of 
fundamental constants based on a simple non-relativistic estimation of 
energy intervals for atomic and molecular transitions 
\cite{savedoff, thompson} (see Table~\ref{t:scal}). Importance of 
the {\em relativistic corrections} was first emphasized in \cite{prestage} 
(studying the hyperfine intervals in different atoms) and later was explored 
in \cite{dzuba} for applications to other transitions realized in most 
accurate microwave and optical measurements. Recently relativistic atomic 
calculations were intensively used for a study of astrophysical data 
in \cite{webb,dzuba}.

\begin{table}[t]
\caption{Scaling behavior of energy intervals as functions of the 
fundamental constants. $Ry$ stands for the Rydberg constant, $\mu$ is the 
nuclear magnetic moment. The references are related to papers where use of 
this scaling behavior in a search for the variations was pointed out. 
Importance of the relativistic corrections for the hyperfine structure was 
first emphasized in \protect\cite{prestage}, while for other atomic 
transitions it was discussed in \protect\cite{dzuba}. \label{t:scal}}
\begin{center}
\begin{tabular}{|c|c|c|c|}
\hline
\multicolumn{2}{|c|}{Transition} & Energy scaling & Refs.\\
\hline
Ato&Gross structure & $ {\rm Ry} $ & \protect\cite{savedoff}\\
\cline{2-4}
mic&Fine structure & $\alpha^2{\rm Ry} $ & \protect\cite{savedoff}\\
\cline{2-4}
&Hyperfine structure & $\alpha^2(\mu/\mu_B){\rm Ry} $ 
& \protect\cite{savedoff}\\
\hline
Mole&Electronic structure & ${\rm Ry} $ & \protect\cite{thompson}\\
\cline{2-4}
cular&Vibration structure & $(m_e/m_p)^{1/2}{\rm Ry} $ 
& \protect\cite{thompson}\\
\cline{2-4}
&Rotational structure & $(m_e/m_p){\rm Ry} $ & \protect\cite{thompson}\\
\hline
\multicolumn{2}{|c|}{Relativistic corrections }
& Extra factor of $\alpha^2$ & \protect\cite{prestage,dzuba}\\
\hline
\end{tabular}
\end{center}
\end{table}

We have to emphasize that a study of atomic and molecular transitions 
in contrast to nuclear energy levels allows a reliable interpretation of the 
results in terms of the fundamental constants. Most of laboratory 
transitions and an essential part of astrophysical ones involve $s$ levels, 
which have a wave function non-vanishing in the vicinity of the nucleus. 
The operators related to relativistic corrections are singular at short 
distances and the relativistic corrections involve $(Z\alpha)^2$ rather than 
$\alpha^2$, even for neutral atoms and  singly charged 
ions \cite{casimir,dzuba}. For heavy nuclei with a large value of the 
nuclear charge $Z$ such relativistic effects  significantly shift the 
non-relativistic result \cite{dzuba}.

As an example of the application of atomic and molecular transitions, we 
collect in Table~\ref{t:lim2} most recent astrophysical results on the 
variation of fundamental constants. The astrophysical results used to be 
originally presented in terms of a variation of a constant at a given value 
of the red shift $z$. We present them in terms of a variation rate assuming 
a linear time dependence in order to be able to compare the astrophysical 
data with laboratory results which are discussed in the following sections. 
Since the linear drift is an open question, one should consider the data in 
Table~\ref{t:lim2} not as a rigorous result, but rather as an illustration.

\begin{table}[t]
\caption{Most recent astrophysical limits for the variation of fundamental 
constants. Comments: $a$ -- based on relativistic corrections; $b$ -- related 
to H$_2$ molecular spectroscopy; $c$ -- originates from a comparison of 
hydrogen HFS interval with hydrogen molecular spectroscopy.\label{t:lim2}}
\vspace{0.2cm}
\begin{center}
\begin{tabular}{|c|c|c|c|
}
\hline
Value & Variation & Reference & Comment\\
\hline
$\partial \ln\alpha/\partial t $ & $(-0.5\pm0.1)\times10^{-15}$ yr$^{-1}$ & \protect\cite{webb}
& a \\
$\partial \ln(m_p/m_e)/\partial t $ & $(5\pm 3)\times 10^{-15}$ yr$^{-1}$ & \protect\cite{ivanchik} & b \\
$\partial \ln(\alpha^2g_p)/\partial t $ & $(-0.3\pm 0.6)\times 10^{-15}$ yr$^{-1}$ & \protect\cite{murphy} & c \\
\hline
\end{tabular}
\end{center}
\end{table}

\section{Hyperfine structure and variation of nuclear magnetic 
moments\label{s:hfs}}

Successes of high resolution spectroscopy (except last few years) were mostly
related to precision measurements of hyperfine structure in neutral atoms 
(cesium, rubidium, hydrogen etc.) or single charged ions (mecrury, ytterbium 
etc.). Even today, despite a dramatic progress with optical transitions which 
presents a revolution in frequency metrology, the most accurate comparison of 
two frequencies is related to a comparison of hyperfine intervals in neutral 
cesium and rubidium \cite{salomon}. It is this experiment that can 
potentially provide us with the most strong laboratory limit on a variation of 
one of fundamental constants.

However, one question remains to be solved: how to express all nuclear 
magnetic moments in terms of a few basic constants. At least for an odd 
nucleus this problem can be solved with the help of so called 
{\em Schmidt model} (see for detail \cite{sgk_cjp}.

The model explains a value of the magnetic moment of a nucleus with an 
odd number of nucleons as a result of the spin and orbit motion of a single 
nucleon, while the others are coupled and do not contribute. This 
approximation is reasonable when the coupled nucleons form closed shells. If 
the shells are not closed, the corrections could be quite large and the model 
is far from perfect as it is clear from Table~\ref{t:hfs} where we present 
some data related to atoms of interest and obtained from microwave 
spectroscopy.

\begin{table}[t]
\caption{Hyperfine splitting and magnetic moment of some alkali atoms. Here: 
$\,u_S$ is the Schmidt value of the nuclear magnetic moment $\mu$ while 
$\mu_N$ is the nuclear magneton. The uncertainty of the calculation in 
\protect\cite{prestage,casimir} is estimated by comparing less accurate 
general results on cesium and mercury in \protect\cite{prestage,casimir} 
with a more accurate calculation in \protect\cite{dzuba}. The actual values 
of the nuclear magnetic moments are taken from \protect\cite{firestone}.
A sensitivity to $\alpha$ variation is defined as 
$\kappa=\partial \ln\big(F_{\rm rel}(\alpha)\big)/\partial \ln\alpha$.
\label{t:hfs}}
\vspace{0.2cm}
\begin{center}
\begin{tabular}{ |c|c|c|c|c|c|}
\hline
$Z$ &Atom & Schmidt value   & Actual value  & Relativistic        & $\kappa$ \\
    &     & for $\mu$   & for $\mu$ & factor              & \\
    &     & ($\mu_S/\mu_N$) & ($\mu/\mu_S$) & $F_{\rm rel}(\alpha)$ &  \\
\hline
1&H              & $g_p/2$        & 1.00 & 1.00 & 0.00\\
4&$^9$Be$^+$     & $g_n/2$        & 0.62 & 1.00 & 0.00\\
37&$^{85}$Rb   & $5/14(8-g_p)$ & 1.57 & 1.15, \cite{prestage,casimir} & 0.30(6)\\
  &$^{87}$Rb      & $g_p/2+1$      & 0.74 & 1.15, \cite{prestage,casimir} & 0.30(6) \\
55&$^{133}$Cs     & $7/18(10-g_p)$ & 1.50 & 1.39, \protect\cite{dzuba} & 0.83 \\
70&$^{171}$Yb$^+$ & $-g_n/6$       & 0.77 & 1.78, \cite{prestage,casimir} & 1.42(15)\\
80&$^{199}$Hg$^+$ & $-g_n/6$       & 0.80 & 2.26, \protect\cite{dzuba} & 2.30 \\
\hline
\end{tabular}
\end{center}
\end{table}

The most accurate up-to-now frequency comparison is related to the hyperfine 
intervals of cesium-133 and rubidium-87 in their ground states \cite{salomon}, 
however, there is a problem of a proper interpretation of the result in terms 
of the fundamental constants due to significant corrections to the Schmidt 
values. 

We have to note that in general the corrections to the Schmidt model are 
comparable to relativistic corrections caused by atomic effects. The 
relativistic corrections expressed in terms of a correcting factor to the 
non-relativistic result
\beq
\Delta E = \Delta E({\rm non\!-\!rel})\times F_{\rm rel}(\alpha)
\eeq
are also summarized in Table~\ref{t:hfs}.

However, it is not very clear which nucleus is good for a search, the one 
with small corrections to the Schmidt model or the one with the large 
corrections. Indeed, in the case of small corrections the interpretation 
is easier. However, as one can see from Sect.~\ref{s:corr}, in some models 
the $g$ factor of the nucleon should vary relatively slow. In the case of 
chiral QCD (a limit of zero masses for the current quarks) the nucleon $g$ 
factors are in principle completely determined. When one takes into account 
the current masses of the $u$ and $d$ quarks, the corrections to the nucleon 
magnetic moment and the nucleon mass are negligible. However, that is not the 
case for the current mass of the $s$ quark, which significantly affects 
the nucleon mass, but not the magnetic moment. In other words, the nuclear 
magneton ($e\hbar/2m_p$) is shifted with the correction to $m_p$, while the 
magnetic moments of proton and neutron are not. Since the $g$ factor is a 
ratio of the nucleon magnetic moment and the nuclear magneton, the $g$ 
factors of the proton and the neutron are to be shifted coherently with 
their ratio not affected. The main property of QCD is that in the chiral 
limit the dimensional value depends on a parameter $\Lambda_{\rm QCD}$, while 
dimensionless ones are independent and fixed. If we accept ideas 
of \cite{calmet} that the variations are somehow caused by variations 
of $\Lambda_{\rm QCD}$, a variation of the $g$ factors has to be 
relatively slow and in principle it may happen that nuclear corrections 
vary faster than $g$ factors. 

We also need to mention a correlation between a relative value of the correction to the Schmidt model and importance of the orbital contribution to the nuclear magnetic moment. As one can see from Table~\ref{t:hfs} the biggest corrections to the Schmidt model ($^{87}$Rb and $^{133}$Cs) related to a distructive interference between the spin and orbit contributions. The partial cancelation between these two terms which differently depend on fundamental constants offers  hyperfine intervals which are potentially the most sensitive to the variations of the constants. In all other atoms important for frequency standards in Table~\ref{t:hfs} the orbital contribution either small or vanish at all and since the $g_p$ and $g_n$ can vary coherently this intervals could be not quite sensitive to a variation of the QCD parameter $\Lambda_{\rm QCD}$.
In such a situation it is important to study 
both the magnetic moments with small corrections to the Schmidt model and 
those with the large corrections. 

The most accurate microwave measurements are related to hyperfine intervals 
in hydrogen \cite{hydrogen1s}, ytterbium ion \cite{warrington}, 
rubidium \cite{salomon} and cesium. A value of the latter is accepted by 
definition but one has to provide a realization of this definition which 
involves experimental uncertainties. The experiment with four mentioned hyperfine intervals were performed several times and 
provide us with some limits on the variation of their frequency. Most of 
experiments were carried out as {\em absolute measurements}, i.e. as a 
comparison of a frequency under study with the cesium hyperfine interval. 
A few more measurements were performed as {\em relative measurements} with 
a direct comparison of two different frequencies. Work on some experiments 
is not yet completed and we can hope that their results will be repeated with 
higher accuracy. There are also some experiments (see \cite{sgk_cjp} for 
detail) which were done with high accuracy only once and unfortunately 
there are no plans to repeat them in close future: on 
beryllium \cite{bollinger} and mercury \cite{berkland}.

The Schmidt model also predicts that some nuclear magnetic moments have to 
be quite {\em small} because of a significant cancellation between the 
proton spin and orbit contributions, in particular for nuclei of $1/2^-$ 
and $3/2^+$ (see \cite{sgk_cjp} for detail). Those magnetic moments as well 
as small magnetic moments of even nuclei can be used to look for a variation 
of the fundamental constants since an enhancement may arise in them because 
of a substantial cancellation between contributions with different dependence 
on fundamental constants. Extending our interest to even and radioactive 
nuclei, we can note that a cancellation between different contributions 
(proton and neutron spins, proton orbit) can lead to a dramatically small 
value. For instance, the nuclear magnetic moment of 
${}^{198}$Tl ($T_{1/2}=5.3(5)$~h) is below than a part in $10^3$ of the 
nuclear magneton.

We have to note that often two additional neutrons do not change the 
nuclear magnetic moment very much and the magnetic moments for different odd 
isotopes of the same element are about the same. For rubidium-85 ($I = 5/2$) 
and for rubidium-87 ($I = 3/2$) that is not a case and their comparison will 
be free of a variation of the fine structure constant, so that by comparing 
them one can study a variation of $g_p$ only. 

\section{Optical transitions and variation of the fine structure 
constant\label{s:opt}}

Optical transitions are free of the problem of nuclear effects. There are 
two kinds of optical experiments:
\begin{itemize}
\item One can compare an optical frequency with another optical one.
\item One can measure an absolute optical frequency, i.e. compare an optical 
frequency with a hyperfine interval in cesium.
\end{itemize}
The latter suggests a mixed optical--microwave experiment, however, the 
cesium frequency may be excluded from the final interpretation. The 
optical--optical and optical--microwave high precision relative measurements 
are now possible due to the recently developed frequency comb \cite{udem}.

The most accurate measurements were performed with the hydrogen 
\cite{niering} and calcium \cite{udem1} atoms and with the ytterbium 
\cite{stenger} and mercury \cite{udem1,bize} ions. All these results can 
contribute in accurate limits on the variation of the constants if they are 
repeated with high accuracy, since previous measurements were not as accurate 
as the most recent ones. As far as I know, all experiments are in progress 
and new results have to be expected soon. Somewhat less accurate results 
were achieved some time ago for the indium \cite{zanthier} and strontium 
\cite{lea} ions and further progress is still possible.

\begin{table}[t]
\caption{The most accurate optical measurements: results and sensitivity of 
the optical transitions to a time variation of $\alpha$. The sensitivity to 
$\alpha$ variation is defined as 
$\kappa=\partial \ln\big(F_{\rm rel}(\alpha)\big)/\partial \ln\alpha$ and 
presented according to \protect\cite{dzuba}, except for the strontium 
result ($\star$) which is our rough estimate based on \protect\cite{dzuba}.
\label{t:opt}}
\vspace{0.2cm}
\begin{center}
\begin{tabular}{|c|c|c|c|c|}
\hline
$Z$&Atom & Frequency & Fractional & $\kappa$ \\
&& [Hz] &uncertainty & \protect\cite{dzuba}\\
\hline
1&H      & 2\,466\,061\,413\,187\,103(46), \protect\cite{niering}& $2\times 10^{-14}$ & 0.00 \\
20&Ca     & 455\, 986\,240\,494\,158(26), \protect\cite{udem1}& $6\times 10^{-14}$ & 0.03\\
38&Sr$^+$ & 444\,779\,044\,095\,520(100), \protect\cite{lea}& $22\times 10^{-14}$ & $0.25^\star$\\
49&In$^+$ & 1\,267\,402\,452\,899\,920(230), \protect\cite{zanthier}& $18\times 10^{-14}$ & 0.21 \\
70&Yb$^+$ & 688\,358\,979\,309\,312(6), \protect\cite{stenger}& $0.9\times 10^{-14}$ &0.9 \\
80&Hg$^+$ & 1\,064\,721\,609\,899\,143(10), \protect\cite{udem1}& $0.9\times 10^{-14}$ & - 3.18\\
& & 1\,064\,721\,609\,899\,144(14), \protect\cite{bize}& $1.0\times 10^{-14}$ & \\
\hline
\end{tabular}
\end{center}
\end{table}

The claimed accuracy of experiments \cite{niering,udem1,stenger} 
(see Table~\ref{t:opt}) was at the level of a part per $10^{14}$ and 
one may doubt their high precision as well as in that in the case of 
microwave experiments \cite{salomon,hydrogen1s,bollinger,warrington,berkland}. 
In most laboratories the results were reproduced for a number of times 
with a different setup, e.g., with different ion traps. A few experiments 
were independently  performed in several laboratories:
\begin{itemize}
\item the hydrogen hyperfine splitting was numerously measured in 
1970-1980 \cite{hydrogen1s} (see also \cite{sgk_cjp} for detail);
\item cesium standards from different laboratories were numerously compared one to 
another (see e.g. \cite{parker});
\item the ytterbium microwave interval was measured a few times at 
PTB \cite{tamm} and NML \cite{warrington};
\item the calcium optical transition was measured a few times at 
PTB \cite{riehle} and NIST \cite{udem1}.
\end{itemize}
Only one of the four mentioned above quantities is related to optical 
transitions. However, good understanding of a cesium standard is crucially 
important if we like to interpret two absolute measurements of different 
optical transitions realized in different laboratories as an indirect 
measurement of their ratio. There are also other possibilities to 
verify accuracy of spectroscopic experiments.

\begin{figure}
\centerline{
\epsfig{figure=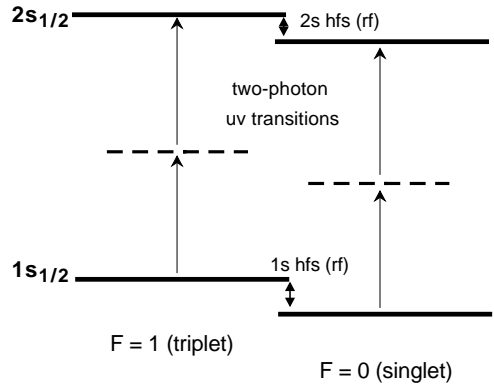,height=2in}}
\caption{The level scheme for an optical measurement of the 2s hyperfine 
structure in the hydrogen atom \cite{fischer}.\label{f:2shfs}}
\end{figure}

A test for the $1s-2s$ transition was recently performed at MPQ \cite{fischer}.
A study of transitions for different hyperfine components and determination 
of their difference offer a possibility to find the hyperfine interval for 
the metastable $2s$ state. The uncertainty of the $2s$ hyperfine interval 
is $6\times10^{-15}$ of the big $1s-2s$ interval (see Fig.~\ref{f:2shfs}). 
The optical result \cite{fischer} for this microwave quantity has an accuracy 
higher than the recent radio frequency result \cite{rothery} and it is in  
good agreement with theory \cite{ki_epj}.

\begin{figure}
\centerline{\epsfig{figure=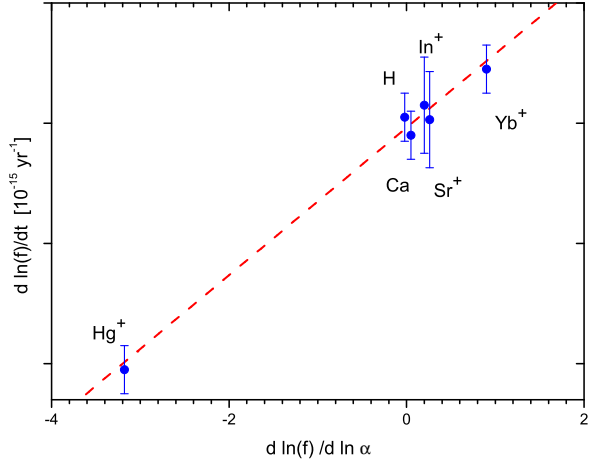,height=3.0in}}
\caption{
The expected structure of the optical data for a possible variation of 
the fundamental constants. A value of 
$\partial F_{\rm rel}(\alpha)/\partial \ln(\alpha)$ 
for each transition has been taken from \protect\cite{dzuba} (see Table 4 for detail).\label{f:anchor}}
\end{figure}

Several transitions for hydrogen, calcium, ytterbium and mercury have been 
accurately measured a few times or monitored for some period. However, 
the monitoring was performed for a relative short period (a few months rather
than a few years) and in the case of a comparison of two measurements 
separated by years, the second wave of the measurements (see 
Table~\ref{t:opt}) was more accurate than the first wave. An exception is a 
recent measurement on mercury \cite{bize}.
After experiments \cite{niering,udem1,stenger} are reproduced with higher 
accuracy, we will obtain four values for a variation of the optical 
frequencies with respect to the cesium microwave frequency. The frequency 
$f$ of an optical transition can be presented in the form
\beq
f = {\rm const} \times Ry \times F_{\rm rel}(\alpha)\;.
\eeq
A possible variation of the value of the Rydberg constant in SI units is 
related to a variation of the cesium hyperfine interval in natural atomic 
units and cannot have  simple interpretation (see Sect.~\ref{s:hfs}). 
However, we can compare a {\rm relative} drift of optical frequencies as a 
function of $\partial F_{\rm rel}(\alpha)/\partial \ln(\alpha)$. The 
expected signature of the variation of the fine structure constant is 
presented in Fig.~\ref{f:anchor}. The dashed line 
\beq
\frac{\partial \ln(f)}{\partial t} = a + b \times \frac{\partial F_{\rm rel}(\alpha)}{\partial \ln(\alpha)}\;,
\eeq
is related to the time variation: 
\beq
a = \frac{\partial \ln(Ry)}{\partial t}
\eeq
of the Rydberg constant in the SI units and 
\beq
b={\partial \ln\alpha}/{\partial t}
\eeq 
of the fine structure constant $\alpha$.

\section{Application of frequency measurements 
to a search for the violation of the Lorentz invariance\label{s:lor}}

A search for a time variation of a ratio of two frequencies can also be 
applied to test the Lorentz invariance. A breakdown of this invariance 
assumes existence of a favorite frame, but we actually have one. That is the 
frame where the microwave background radiation is isotropic. An only question is 
whether there are any effects depending on the velocity $v$ with respect to 
this frame or not. A violation of the Lorentz invariance can, for example,
lead to different summation of two velocities. A transition energy has to be 
of the form:
\begin{eqnarray}\label{Lor:an}
\Delta E &=& c_1\cdot \alpha^2mc^2 \times \left(1+c_2\cdot \frac{v^2}{c^2}\right) + c_3\cdot \alpha^4mc^2\times \left(1+c_4\cdot\frac{v^2}{c^2}\right) + Å\nonumber\\
&=& \Delta E_{v=0} \times \left(1+c_{\rm eff} \cdot \frac{v^2}{c^2} \right)\;.  
\end{eqnarray}

The coefficients $c_2$ and $c_4$ are rather of kinematic origin and it can 
happen that they depend on an atom only slightly. However, if relativistic 
corrections are big enough, an anomalous coefficient $c_{\rm eff}$ definitely 
has to be different. The velocity of the Sun with respect to the frame where 
the microwave background is isotropic is well known \cite{fixsen}. Because of 
the Earth motion with respect to the Sun, the value of $v^2/c^2$ is changing 
periodically with the amplitude of $2.5\times 10^{-7}$. 

The expected amplitude of the annual variation of the value of $v^2/c^2$ in 
Eq. (\ref{Lor:an}) actually does not depend strongly on the assumption 
what frame is the favorite. However, the assumption that the favorite frame 
is the one where the microwave background radiation is isotropic, can 
essentially simplify analysis because of a known phase of the annual 
oscillations.

By now, a number of systems with great relativistic corrections have been 
studied (atomic transitions, nuclear magnetic moments etc). The highest 
accuracy is slightly better than a part per $10^{14}$. However, the biggest 
uncertainty is rather systematic than statistical and can be dismissed when 
looking for a periodic oscillation with a period of a sidereal year with a 
known phase. We expect that the limit for $c_{\rm eff}$ at the level of 
a few ppb is feasible.

\section{Summary\label{s:sum}}

\subsection{Actual laboratory limits for variation of fundamental constants\label{s:act}}

There are two kinds of comparisons of different frequencies. One is related to 
so-called clock comparisons. In other words, the idea of the experiment is 
to compare two clocks, based on different transitions. The other kind of 
comparison is a direct comparison of two measured frequencies. When a 
relation of a transition frequency and clock frequency is well understood, 
a clock comparison is the same as a comparison of the transition frequencies. 
However, it is often not the case, because there is a number of systematic 
effects responsible for a difference between an unperturbed value of some 
transition and a reference frequency produced by the clock. A well known 
example when they are essentially not the same is a hydrogen mazer. The 
effects due to the wall shift, which are actually time dependent (and that 
is clearly seen from long-term measurements), are essentially bigger than 
a short-term unstability. In contrast to a number of papers 
(see, e.g., \cite{uzan}) we follow \cite{sgk_cjp} and completely exclude 
any clock comparisons from our consideration paying attention to comparisons 
of transition frequencies only. We collect current laboratory limits for 
the variation of frequencies of atomic transitions of gross, fine and 
hyperfine structure in Table~\ref{t:lim1}. 

\begin{table}
\caption{Current laboratory limits on variation of frequency 
$(1/f)\,\vert(\partial f)/(\partial t)\vert$ in SI units 
(i.e. with respect to a hyperfine interval in cesium. Limits marked with 
an asterix ($*$) are obtained by comparing two or more separately published 
measurements, while the ones marked with a star ($\star$) are taken from 
direct monitoring of two frequencies within the same long-term experiment.\label{t:lim1}}
\begin{center}
\begin{tabular}{|c|c|c|c|}
\hline
Atoms&Limit&Ref(s).&Comment\\
\hline
H& $2\times10^{-13}$ yr$^{-1}$ &\protect\cite{hydrogen1s,sgk_cjp}$^*$& Hyperfine\\
Rb&$7\times10^{-16}$ &\protect\cite{marion}$^\star$&structure\\
Yb$^+$&$5\times10^{-14}$ &\protect\cite{fisk,warrington}$^*$&\\
\hline
H&$1\times10^{-13}$ &\protect\cite{udem2,niering}$^*$&Gross \\
Ca&$8\times10^{-14}$ &\protect\cite{schnatz,riehle,udem1}$^*$& structure\\
Hg$^+$&$7\times10^{-15}$ &\protect\cite{udem1,bize}$^*$&\\
\hline
Mg&$3\times10^{-13}$ &\protect\cite{godone}$^\star$& Fine structure\\
\hline
\end{tabular}
\end{center}
\end{table}

Dependence of the frequencies on the fundamental constants is discussed 
above and one can interpret the data from Table~\ref{t:lim1} in terms of 
the fundamental constants. There are various ways to achieve a pure optical 
result for the variation. One can combine two gross structure transitions and 
take advantage of different values of relativistic corrections. Another 
possibility is to combine a gross structure transition and a fine structure 
one and to use their different non-relativitic behavior as a function of 
fundamental constants. In both cases the hyperfine interval of cesium can 
be used as a reference line and its effects will be canceled in the final 
result assuming a variation with a cosmological scale. We note, however, 
that a variation of atomic frequencies can in principle be caused by a 
violation of the Lorentz invariance (see, e.g., the previous section). 
In such a case a variation of the frequency ratio should have a fast 
component and a combination of results from a few separate experiments 
requires an additional analysis.
 
The actual limits for the variation of the fundamental constants are 
summarized in Table~\ref{t:lim}. To estimate the variation of $g_p$ we use 
the result \cite{salomon}. However, for all other limits we combine a 
comparison of some probe frequency with the cesium hyperfine interval and 
the result \cite{salomon}. Thus, we exclude cesium from consideration and 
effectively deal with the variation of the probe frequency with respect to 
the rubidium hyperfine interval. The latter is understood better than cesium 
within the Schmidt model and thus provides us with more reliable data.

We note that recently a number of optical transitions were measured only 
once and data coming in 2003 and 2004 are expected to improve essentially a limit for 
a variation of the fine structure constant and thus of $\mu_p/\mu_e$ and 
$m_p/m_e$.

\begin{table}[t]
\caption{Actual laboratory limits for a variation of the fundamental 
constants (see \protect\cite{sgk_cjp,sgk_60,sgk_tam} for detail). All 
results but a limit for $g_p$ are model independent. The $g_p$ result is 
based on the Schmidt model. \label{t:lim}}
\vspace{0.2cm}
\begin{center}
\begin{tabular}{|c|c|}
\hline
Fundamental & Limit for\\
constant & variation rate\\ \hline
$\alpha$ & $1\times 10^{-14}$ yr$^{-1}$ \\ \hline
$\alpha^2 \mu_p/\mu_e$ & $6\times 10^{-14}$ yr$^{-1}$ \\ \hline
$\alpha^{-3} \mu_p/\mu_e$ & $7\times 10^{-15}$ yr$^{-1}$ \\ \hline
$\alpha^2 \mu_n/\mu_e$ & $8\times 10^{-14}$ yr$^{-1}$ \\ \hline
$\mu_p/\mu_e$ & $2\times 10^{-14}$ yr$^{-1}$ \\ \hline
$\mu_n/\mu_e$ & $6\times 10^{-14}$ yr$^{-1}$ \\ \hline
$g_n/g_p$ & $5\times 10^{-14}$ yr$^{-1}$ \\ \hline
$m_e/m_p$ &$2\times 10^{-13}$ yr$^{-1}$ \\ \hline
$g_p$ & $4\times 10^{-16}$ yr$^{-1}$ \\ \hline
\end{tabular}
\end{center}
\end{table}

\subsection{Comparison of laboratory searches to others\label{s:comp}}

Advantages and disadvantages of different searches for the variation of 
the fundamental constants are summarized in Table~\ref{t:advant} 
(see \cite{sgk_cjp,sgk_60,sgk_tam} for detail). We note that the 
sensitivity of various methods to the separation of space and time 
variations is different. Also different is a possibility to distinguish between an   
oscillation and a linear drift within a cosmological scale. The level of limits retaled to different epochs is quite different, 
however, it is absolutely unclear how the variations should behave with time. 
For instance it may happen that they slow down. Access to different constants and 
reliability of results also differ from one approach to another.

A crucial problem for a reliable interpretation is the involvement of 
strong interactions. In principle, it may be possible to avoid the nuclear 
corrections to the Schmidt model in a study of microwave transitions. 
For example, a limit of a possible variation of the fine structure constant 
could be achieved from a comparison of the hyperfine interval in the ground 
state of single charged ions of ytterbium-171 and mercury-199. We note that 
the Schmidt values for these nuclei are the same. The actual values are 
somewhat below the Schmidt values (by approximately 20\%), but they 
are also approximately the same (within 4\%) for both elements (see 
Table~\ref{t:hfs}). If that is a systematic effect, the corrections to the 
Schmidt model for ytterbium and mercury can cancel each other and a variation of the frequency is to be completely determined by a possible variation of $\alpha$. Unfortunately,
there is no satisfactory understanding if the correction is a 
systematic effect, or it accidentally has about the same value for 
two elements. 

\begin{table}[t]
\caption{
Comparison of different kinds of a search for the variation of the 
fundamental constants (see \protect\cite{sgk_cjp,sgk_60,sgk_tam} for detail). 
\label{t:advant}}
\vspace{0.2cm}
\begin{center}
\begin{tabular}{|c|c|c|}

\hline
&Geochemistry & Astrophysics  \\
\hline
Drift or oscillation & $\Delta t \sim 10^9$ yr & $\Delta t \sim 10^9-10^{10}$ yr \\ \hline
Space variations & $\Delta l \simeq 10^9~c\times$yr & $\Delta l \simeq 
\bigl(10^9-10^{10}\bigr)~c\times$yr \\ \hline
Level of limits & $10^{-17}$ yr$^{-1}$ & $10^{-15}$ yr$^{-1}$  \\ \hline
Present results & Negative & Positive ($\alpha$)  \\ \hline
Variation of $\alpha$ & Not reliable & Accessible \\ \hline
Variation of $m_e/m_p$ & Not accessible & Accessible \\ \hline
Variation of $g_p$ & Not accessible & Accessible  \\ \hline
Variation of $g_n$ & Not accessible & Not accessible  \\ \hline
Strong interactions & Not sensitive & Not sensitive \\ 
\hline
\hline
&Laboratory & Laboratory  \\
&&(optics) \\
\hline
Drift or oscillation  & $\Delta t \sim 1-30$ yr & $\Delta t \sim 1-10$ yr \\ \hline
Space variations  & 0 & 0 \\ \hline
Level of limits & $10^{-15}$ yr$^{-1}$ & $10^{-14}$ yr$^{-1}$ \\ \hline
Present results & Negative & Negative \\ \hline
Variation of $\alpha$  & Accessible & Accessible \\ \hline
Variation of $m_e/m_p$  & Accessible & Not accessible \\ \hline
Variation of $g_p$  & Accessible & Not accessible \\ \hline
Variation of $g_n$  & Accessible & Not accessible \\ \hline
Strong interactions & Sensitive & Not sensitive \\ \hline
\end{tabular}
\end{center}
\end{table}

A clear advantage of the optical measurements is their reliable 
interpretation for each particular measurement and a possibility to 
compare results of different experiments.

Because of different sensitivity to various effects and different possible 
hierarchy which can take place because of different scenaria, we believe 
that it is worth trying to search for possible variations using different
methods and a wide variety of studied quantities. A study of optical transitions 
is one of those, which have clear advantages and disadvantages discussed 
above, and we hope that progress in frequency metrology in the optical and 
ultraviolet domain will soon deliver us new accurate results.

Recent progress in optical spectroscopy allows us to hope that new data on 
optical frequency measurements will appear within one or two years and the 
optical limits on a possible variation of the fine structure constant 
will be improved.

\section*{Acknowledgments}

The author is grateful to Z. Berezhiani, V. Flambaum, H. Fritzsch, T. W. H\"ansch, J. L. Hall, L. Hollberg, M. Kramer, W. Marciano, M. Murphy, A. Nevsky, L. B. Okun, E. Peik, T. Udem, D. A. Varshalovich, M. J. Wouters and J. Ye for useful and stimulating discussions.  The consideration in Sect.~\ref{s:lor} was stimulated in part by a discussion with W. Phillips at HYPER symposium who pointed out that for the energy of the $s$ states only effects quadratic in velocity could be important. The work was supported in part by RFBR under grants \#\# 00-02-16718, 02-02-07027, 03-02-16843.

\end{document}